\documentstyle[twoside,fleqn,espcrc2]{article}
\newcommand{\func}[2]{#1\!\left(\!#2\!\right)}
\newcommand{\txtfrac}[2]{\phantom{1\!\!\!}^{#1}\!\!/\!_{#2}}
\newcommand{\etal}[1]{{\it et~al.}~\cite{#1}}
\newcommand{\two}[2]{\begin{tabular}{l} $+{#1}$ \\ $-{#2}$ \end{tabular}}
\newcommand{\val}[4]{${#1}^{+{#2}}_{-{#3}}{({#4})}$}
\newcommand{\fr}{-\frac{8}{5}}
\newcommand{\ph}{\phantom{-\frac{8}{5}}}

\title{A Study of the Static-Light $B_B^{}$ Parameter}

\author{Joseph Christensen\thanks{Presented by J.\ Christensen at
         Lattice '96, St.\ Louis.}, 
        Terrence Draper 
    and Craig McNeile\thanks{Currently at Department of Physics,
         University of Utah, Salt Lake City, UT 84112. } 
        \address{Department of Physics and Astronomy, University of
         Kentucky, Lexington, KY 40506}
        \thanks{This work is supported in part by the U.S. Department
         of Energy under grant numbers DE-FG05-84ER40154 and
         DE-FC02-91ER75661, and by the University of Kentucky Center
         for Computational Sciences.  The computations were carried
         out at NERSC.}}

\begin{document}

\begin{abstract}
We calculate the $B_B^{}$ parameter, relevant for
$\overline{B}^0_{}${\bf --}$B^0_{}$ mixing, from a lattice gauge theory
simulation using the static approximation for the heavy quark and the
Wilson action for the light quark and gauge fields.  Improved sources,
produced by an optimized variational technique, {\sc most}, reduce
statistical errors and minimize excited-state contamination of the
ground-state signal.  Renormalization of four-fermion operator
coefficients, using the Lepage-Mackenzie procedure for estimating
typical momentum scales, is linearized to reduce order $\alpha_s^2$
uncertainties.
\end{abstract}

\maketitle


\section{$B_B^{}$ Parameter}

Since the lattice static effective theory has fewer symmetries than the
full continuum theory, when calculating the static-light $B_B^{}$
parameter
\begin{equation}
B_B^{} = \frac{ \left< \overline{B}_0
                \left| {\cal O}_L^{{\rm full}} 
                \right| B_0 \right>_{\phantom{vs}}}
              { \left< \overline{B}_0
                \left| {\cal O}_L^{{\rm full}} \right| B_0 \right>_{vs} }
\end{equation}
operators besides ${\cal O}_L^{{\rm latt}}$ must be included.  These
correspond to the following full-theory fermion operators (see
Flynn~\etal{Flynn91a}):
\begin{eqnarray}
{\cal O}_L
& \!\! = \!\!
    & \frac{1}{2}
      \left( \overline{b} \gamma^\mu P_L q \right)
      \left( \overline{b} \gamma_\mu P_L q \right)
      \nonumber \\
{\cal O}_R
& \!\! = \!\! 
    & \frac{1}{2}
      \left( \overline{b} \gamma^\mu P_R q \right)
      \left( \overline{b} \gamma_\mu P_R q \right)
      \nonumber \\ 
{\cal O}_N
& \!\! = \!\! 
    & \frac{1}{2} \left[
      \left( \overline{b} \gamma^\mu P_L q \right)
      \left( \overline{b} \gamma_\mu P_R q \right)
    + \right. \nonumber \\ & & \,\, \left.
    + \left( \overline{b} \gamma^\mu P_R q \right)
      \left( \overline{b} \gamma_\mu P_L q \right)
    + \right. \nonumber \\ & & \,\, \left.
    + 2 \left( \overline{b} P_L q \right)
        \left( \overline{b} P_R q \right)
    + 2 \left( \overline{b} P_R q \right)
        \left( \overline{b} P_L q \right) \right]
      \nonumber \\
{\cal O}_S
& \!\! = \!\!
    & \frac{1}{2}
      \left( \overline{b} P_L q \right)
      \left( \overline{b} P_L q \right)
      \label{eq:ops}
\end{eqnarray}
${\cal O}_S$ is generated at order $\alpha_s$ in the continuum due to
the mass of the heavy quark.  ${\cal O}_R$ and ${\cal O}_N$ are
generated at order $\alpha_s$ from the chiral symmetry breaking Wilson
mass term.  The lattice calculation of the static-light $B_B^{}$ uses
the ratio of two- and three-point hadronic correlation functions.
\begin{eqnarray}
\func{B}{t_1,t_2}
& \!\! = \!\! 
    & \frac{ \func{C_3^{}}{t_1,t_2} }
           { \txtfrac{8}{3}
             \func{C_2}{t_1} 
             \func{C_2}{t_2} }
    \ \stackrel{ \left| t_i \right| \,\gg 1 }
               { -\!\!\!-\!\!\!-\!\!\!\longrightarrow }\ 
      B_B
\label{eq:B_to_B}
\end{eqnarray}
where the required correlation functions are
\begin{eqnarray*}
\lefteqn{\func{C_3^{}}{t_1,t_2}} \\
& \!\! = \!\! 
    & \sum_{\vec{x}_1,\vec{x}_2}
      \left< 0 \left| {\cal T} \left(
      \func{\chi}{t_1,\vec{x}_1\,}
      \overline{b} ( 0,\vec{0} )
      \Gamma_{\!\!I}^{}
      q(0,\vec{0})
      \right. \right. \right. \nonumber \\ & & \left. \left. \left.
      \phantom{00000}
      \overline{b}(0,\vec{0})
      \Gamma_{\!\!J}^{}
      q(0,\vec{0})
      \func{\chi}{t_2,\vec{x}_2\,}
      \right) \right| 0 \right> \\
\lefteqn{\func{C_2}{t_1}} \\
& \!\! = \!\! 
& \!  \sum_{\vec{x}_1}
      \left< 0 \left| {\cal T} \left(
      \func{\chi}{t_1,\vec{x}_1\,}
      \overline{b}(0,\vec{0})
      \gamma_4^{} \gamma_5^{}
      q(0,\vec{0})
      \right) \right| 0 \right>
\end{eqnarray*}
The three-point function has a fermion operator inserted at the
spacetime origin, between two external $B$-meson interpolating fields.
The times are restricted to the range $t_2>0>t_1$.  The gamma matrices,
$\Gamma_{\!\!I\,}^{}$ and $\Gamma_{\!\!J\,}^{}$ define the type of four
fermion operator (see equation~\ref{eq:ops}).  A spatially extended
$B$-meson operator
\begin{equation}
\func{\chi}{t,\vec{x}\,}
  =   \sum_{\vec{r}}
      \func{f}{\vec{r}\,}
      \func{\overline{q}}{t,\vec{x}+\vec{r}\,}
      \gamma_5^{}
      \func{b}{t,\vec{x}\,}
\end{equation}
is used, where $f$ is a smearing function produced by {\sc
most}~\cite{Draper94a} for our static $f_B$ study.


\section{Scale Formulation}

Using the integrand of the one-loop perturbative contribution from the
coefficients as a weighting function, as per Lepage and
Mackenzie~\cite{Lepage93}, a ``typical'' momentum scale can be found
(Table~\ref{t:qstar}).
\begin{table}
\begin{center}
\begin{tabular}{ccccc} \hline \hline
& \multicolumn{1}{c}{$\left<{\cal O}_L\right>$}
& \multicolumn{1}{c}{$\left<{\cal O}_L^{{\rm full}}\right>$}
& \multicolumn{1}{c}{$\left<A_\mu\right>$}
& \multicolumn{1}{c}{$B_B$} \\ \hline
$q^{\star}_i a$ 
& $2.01$ 
& $2.15$   
& $2.18$   
& $0.82$ \\ \hline \hline
\end{tabular}
\caption{\sl ``Typical'' operator scales; using $\beta$=6.0 and $r$=1.}
\label{t:qstar}
\end{center}
\vspace{-\bigskipamount}
\vspace{-\bigskipamount}
\end{table}
Our value for the scale relevant for $\left<A_\mu\right>$ agrees with
that found by Hern\'andez and Hill~\cite{Hernandez94a}.  This is the
scale which we claim is relevant for this calculation as well.  We
notice that the scale found for $B_B$ is singularly different than the
others and claim that each of the other matrix elements is describing
physics at essentially the same scale.  However, when a ratio is
considered, the integrands should cancel, but the scale should not.
Since the other values are similar, we choose 2.18, as it has been used
for the $f_B^{}$ study.
%


\section{Calculation of the Coefficients}

The coefficients of the operators are
calculated~\cite{Flynn91a,Gimenez93a,Buchalla96a} by renormalization
group techniques.
\begin{eqnarray}
\func{B_B^{{\rm full}}}{m_b} \!\!\!\!\! \!\!\!\!\! \!
& \!\!\!\!\! & \: = \sum_{i=L,R,N,S}  Z_{B_i} B_i^l \\
Z_{B_L}
& \!\! = \!\!
    & \left( \frac{\func{\alpha_s^c}{m_b}}
                  {\func{\alpha_s^c}{\mu}} 
             \right)^{\left(p_{0,L}^c-2p_{0,f}^c\right)}
      \left[ 1 \phantom{\frac{1}{1}}
             \nonumber \right. \\ & & \left. 
             \ \ \ \ 
           + \frac{ \func{\alpha_s^c}{m_b}
                  - \func{\alpha_s^c}{\mu}}{4\pi} 
             \left( p_{1,L}^c -2 p_{1,f}^c \right)
             \right. \nonumber \\ & & \left.
             \ \ \ \ 
           + \frac{\func{\alpha_s^c}{m_b}}{3 \pi}
             \left(\frac{-13}{2}\right)
           + \frac{\func{\alpha_s^l}{\mu}}{3 \pi}
             \left(11.56\right)
             \right]
      \nonumber \\
Z_{B_R} 
& \!\! = \!\!
    & \left( \frac{\func{\alpha_s^c}{m_b}}
                  {\func{\alpha_s^c}{\mu}} 
             \right)^{\left(p_{0,L}^c-2p_{0,f}^c\right)}
      \frac{\func{\alpha_s^l}{\mu}}{3 \pi} \left( -1.205 \right)
      \nonumber \\
Z_{B_N} 
& \!\! = \!\!
    & \left( \frac{\func{\alpha_s^c}{m_b}}
                  {\func{\alpha_s^c}{\mu}} 
             \right)^{\left(p_{0,L}^c-2p_{0,f}^c\right)}
      \frac{\func{\alpha_s^l}{\mu}}{3 \pi} \left( -10.83 \right)
      \nonumber \\
Z_{B_S}
& \!\! = \!\!
    & \left( \frac{\func{\alpha_s^c}{m_b}}
                  {\func{\alpha_s^c}{\mu}} 
             \right)^{\left(p_{0,S}^c-2p_{0,f}^c\right)}
      \frac{\func{\alpha_s^c}{m_b}}{3 \pi} \left( -6 \right)
      \nonumber
\end{eqnarray}
where $p_{0,i}^{} = \frac{\gamma_{0,i}^{}}{2 b_0^{}}$, and $p_{1,i}^{}
= p_{0,i}^{} \left( \frac{\gamma_{1,i}^{}}{\gamma_{0,i}^{}} -
\frac{b_1^{}}{b_0^{}} \right)$.  For $\alpha_s^l$, we use
$\alpha_V^{}(q^\star)$=0.18 \cite{Lepage93}.

The statistical uncertainties for the coefficients are listed in
Table~\ref{tb:Bcoefferr}.  There is a systematic error due to the
linearization of the coefficients which is not listed.  See
reference~\cite{Christensen96} for complete details.


\section{Results of Simulation} 

The raw lattice $B$ parameters for the operators which appear in the
lattice-continuum matching are determined from a Monte Carlo
calculation of equation~\ref{eq:B_to_B} and listed in
Table~\ref{tb:Braw}.  Table~\ref{tb:Bval} lists the linear combination
$B_{{\cal O}_{L}^{\rm full}}=B_{B}$ as a function of $\kappa$ and
extrapolated to $\kappa_{c}$ using fully-linearized tadpole-improved
coefficients.  For both tables, the first errors are statistical
(bootstrap) and the second are systematic due to choice of fit range.
\begin{table*}
\begin{center}
\begin{tabular}{rccccc} \hline \hline
 & $q^{\star}a$ & $a^{-1}$ &  $m_{b}$  & $\Lambda_c^{(5)}$ & All \\
 &    2.18      & 2.1\,GeV & 4.33\,GeV &     0.175\,GeV    &     \\
\hline
\multicolumn{6}{c}{$Z_{B_L}=1.070$} \\
\hline
10\%    & \two{0.002}{0.002}   & \two{0.003}{0.004} & 
          \two{0.003}{0.003}   & \two{0.0008}{0.0005} & 
          \two{0.004}{0.005} \\
20\%    & \two{0.005}{0.003}   & \two{0.006}{0.009} & 
          \two{0.006}{0.005}   & \two{0.0019}{0.0009} & 
          \two{0.008}{0.009} \\
\hline
\multicolumn{6}{c}{$Z_{B_R}=-0.0225$} \\
\hline
10\%    & \two{0.0005}{0.0006} & & & & \two{0.0005}{0.0006} \\
20\%    & \two{0.0009}{0.0015} & & & & \two{0.0009}{0.0015} \\
\hline
\multicolumn{6}{c}{$Z_{B_N}=-0.202$} \\
\hline
10\%    & \two{0.005}{0.006}   & & & & \two{0.005}{0.006} \\
20\%    & \two{0.008}{0.012}   & & & & \two{0.008}{0.012} \\
\hline
\multicolumn{6}{c}{$Z_{B_S}=-0.137$} \\
\hline
10\%    & & & \two{0.003}{0.003} & \two{0.002}{0.003} 
            & \two{0.003}{0.004} \\
20\%    & & & \two{0.006}{0.005} & \two{0.005}{0.007}
            & \two{0.006}{0.008} \\ \hline \hline
\end{tabular}
\caption{\sl The absolute changes from our preferred values of the
coefficients $Z_{B_L}$, $Z_{B_R}$, $Z_{B_N}$, and $Z_{B_S}$ as the
parameters $q^{*}a$, $a^{-1}$, $m_{b}$, and $\Lambda_c^{(5)}$ are
varied by 10\%, and 20\%, first individually, and then jointly
(``All''), from our preferred input values.  The coefficients are quite
insensitive to the particular choice of input parameter.}
\label{tb:Bcoefferr}
\end{center}
\vfill
\begin{center}
\begin{tabular}{lccccc} 
\hline \hline
 & $\kappa=0.152$ & $\kappa=0.154$ & $\kappa=0.155$ & $\kappa=0.156$ 
 & $\kappa_c=0.157$    \\
\hline
$\ph B_{L}$        & \val{1.01}{2}{2}{1} & \val{1.02}{2}{2}{1} & 
\val{1.02}{3}{2}{1} & \val{1.03}{3}{3}{2} & \val{1.03}{3}{3}{2} \\
$\ph B_{R}$        & \val{0.96}{1}{1}{1} & \val{0.96}{1}{2}{2} & 
\val{0.95}{2}{2}{2} & \val{0.95}{2}{3}{2} & \val{0.95}{2}{3}{2} \\
$\ph B_{N}$        & \val{0.97}{2}{2}{3} & \val{0.96}{2}{2}{4} & 
\val{0.96}{2}{2}{4} & \val{0.96}{3}{2}{4} & \val{0.95}{3}{2}{5} \\
$\fr B_{S}$        & \val{1.00}{2}{1}{2} & \val{1.00}{2}{2}{2} & 
\val{1.00}{2}{2}{3} & \val{1.01}{3}{3}{3} & \val{1.01}{3}{3}{3} \\
\hline \hline
\end{tabular}
\caption{\sl The raw lattice values for the various $B_i$ parameters:
${\cal O}_{S}$ has a vacuum-saturation value different from that of
${\cal O}_{L}$; with a normalization in which the raw $B_i$ have a
common denominator equal to the vacuum-saturated value of ${\cal
O}_{L}$, $\left(-\frac{8}{5}B_{S}\right)$ would identically equal 1.0
if vacuum-saturation were exact.}
\label{tb:Braw}
\end{center}
\vfill
\begin{center}
\begin{tabular}{lccccc} 
\hline \hline
 & $\kappa=0.152$ & $\kappa=0.154$ & $\kappa=0.155$ & $\kappa=0.156$ 
 & $\kappa_c=0.157$    \\
\hline
$B_{B}(m_{b})$ & \val{0.95}{2}{2}{1} & \val{0.96}{3}{2}{2} 
               & \val{0.96}{3}{3}{2} & \val{0.98}{4}{4}{2} 
               & \val{0.98}{4}{4}{3} \\
\hline \hline
\end{tabular}
\caption{\sl $B_B(m_b)$ is calculated by combining the raw $B_i$
parameters with the appropriate coefficients.}
\label{tb:Bval}
\end{center}
\end{table*}
%
%


We find $B_B(m_b)$=\val{0.98}{4}{4}{3}$^{+1}_{-2}$ as our calculated
value.  The first two errors are as mentioned above.  The final error
is due to the statistical uncertainties in the coefficients.  If we run
to a scale of $\mu$=2\,GeV, with $n_{\!f}$=4, using
\begin{equation}
\func{B_B^{}}{\mu} = \left( \frac{\func{\alpha_s^c}{\mu}}
                                 {\func{\alpha_s^c}{m_b}} 
                            \right)^{\left(p_{0,L}^c-2p_{0,f}^c\right)}
                     \func{B_B^{}}{m_b}
\end{equation}
we find $B_B(2\,{\rm GeV})$=$1.05(4)$.  When we convert $B_B(m_b)$ to a
RG invariant quantity using
\begin{equation}
 \hat{B}_B^{}      = \left( \func{\alpha_s^c}{m_b}  
                            \right)^{-\left( p_{0,L}^c
                                           - 2 p_{0,f}^c \right)}
                     \func{B_B^{}}{m_b}
\end{equation}
with 4 flavors, we find $\hat{B}_B$=$1.36(6)$.  With 5 flavors, we find
$\hat{B}_B$=$1.40(6)$.


%
\begin{table*}[t]
\begin{center}
\begin{tabular}{lcccl|cclll} \hline \hline
 & & & scale & & \multicolumn{5}{c}{one-loop} \\
Method & ref & $\beta$ &  (GeV)  & $B({\rm scale})$ 
             & $n_f$ & $\Lambda$ & $B(2.0)$ & $B(4.33)$ & $\hat{B}_B$ \\ 
\hline 
Static-Clover & \cite{Ukqcd95a} & 6.2 & $m_b$=5.0 & 0.69(4) 
  & 5 & 130 & - & - & {\bf 1.02(6)} \\ 
              &                 & 6.2 & $m_b$=5.0 & 0.69(4) 
  & 4 & 200 & 0.75(4) & 0.70(4) & 0.98(6) \\ 
\hline 
Static-Wilson & this & 6.0 & $m_b$=4.33 & 0.98(4) 
  & 5 & 175 & - & - & 1.40(6) \\
              & work & 6.0 & $m_b$=4.33 & 0.98(4) 
  & 4 & 226 & 1.05(4) & {\bf 0.98(4)} & 1.36(6) \\
\hline
Extrap. Static & \cite{Soni95a} & 5.7-6.3 & $\mu$=2.0 & 1.04(5) 
  & 4 & 200 & {\bf 1.04(5)} & 0.97(5) & 1.36(7)\\
               &                & 5.7-6.3 & $\mu$=2.0 & 1.04(5) 
  & 4 & 226 & {\bf 1.04(5)} & 0.97(5) & 1.34(6)\\
Extrap. Static & \cite{Abada92a} & 6.4 & $\mu$=3.7 & 0.90(5) 
  & 0 & 200 & 0.94(5) & 0.89(5) & 1.21(7) \\
               &                 & 6.4 & $\mu$=3.7 & 0.90(5) 
  & 4 & 200 & 0.95(5) & 0.89(5) & 1.25(7) \\ 
\hline
Wilson-Wilson & \cite{Soni95a} & 5.7-6.3 & $\mu$=2.0 & 0.96(6) 
  & 4 & 200 & {\bf 0.96(6}) & 0.90(6) & 1.25(8)\\
              &                & 5.7-6.3 & $\mu$=2.0 & 0.96(6) 
  & 4 & 226 & {\bf 0.96(6}) & 0.89(6) & 1.24(8)\\
Wilson-Wilson & \cite{Soni95a,Bernard88a} & 6.1 & $\mu$=2.0 & 1.01(15) 
  & 4 & 200 & {\bf 1.01(15)} & 0.94(13) & 1.32(20)\\
              &                           & 6.1 & $\mu$=2.0 & 1.01(15) 
  & 4 & 226 & {\bf 1.01(15)} & 0.94(14) & 1.30(19)\\
Wilson-Wilson &  \cite{Jlqcd95a} & 6.1 & $m_b$=5.0 & 0.895(47) 
  & 0 & 239 & 0.96(5) & 0.90(5) & 1.21(6) \\
              &                  & 6.1 & $m_b$=5.0 & 0.895(47) 
  & 4 & 239 & 0.98(5) & 0.91(5) & 1.25(7) \\
              &                  & 6.1 & $m_b$=5.0 & 0.895(47) 
  & 5 & 183 & - & - & 1.29(7) \\ 
Wilson-Wilson & \cite{Jlqcd95a} & 6.3 & $m_b$=5.0 & 0.840(60) 
  & 0 & 246 & 0.90(6) & 0.85(6) & 1.14(8) \\
              &                 & 6.3 & $m_b$=5.0 & 0.840(60) 
  & 4 & 246 & 0.92(6) & 0.85(6) & 1.17(8) \\
              &                 & 6.3 & $m_b$=5.0 & 0.840(60) 
  & 5 & 189 & - & - & 1.20(9) \\ 
Wilson-Wilson & \cite{Abada92a} & 6.4 & $\mu$=3.7 & 0.86(5) 
  & 0 & 200 & 0.90(5) & 0.85(5) & {\bf 1.16(7)} \\
              &                 & 6.4 & $\mu$=3.7 & 0.86(5) 
  & 4 & 200 & 0.91(5) & 0.85(5) & 1.19(7) \\ 
\hline 
Sum Rule & \cite{Narison94a} &  & $m_b$=4.6 & 1.00(15) 
  & 5 & 175 &    -     &    -     & 1.43(22) \\
         &                   &  & $m_b$=4.6 & 1.00(15) 
  & 4 & 227 & 1.08(16) & 1.00(15) & 1.39(21) \\ 
\hline \hline
\end{tabular}
\caption{\sl The authors' numbers, quoted at the listed scale, have
been scaled to $\mu$=2.0\,GeV and to $m_b$=4.33\,GeV.  If the authors
quote a number which we used or reproduced, it is bold-faced in the
table.}
\label{tb:Bothers}
\end{center}
\vspace{-\bigskipamount}
\end{table*}
%


\section{Comparison to Others}

The simulations using Wilson quarks calculate the $B_{B}$ parameter for
quark masses around charm and extrapolate up to the physical mass,
using a fit model of the form $B = B^{0} + \frac{ B^1 }{M}$.  The value
of $B^{0}$ (``extrapolated static'' in Table~\ref{tb:Bothers}) should
be the same as the static theory.  We scale the authors' numbers to
2.0\,GeV and 4.33\,GeV. The JLQCD collaboration cite their $\Lambda$ as
$n_{\!f}$=0 values.  When Abada {\it et al.}\ quote a $\hat{B}_B$ for
the propagating Wilson quarks, they use $n_{\!f}$=0.  When we scale
these, we list values for both $n_{\!f}$=0 and $n_{\!f}$=4.  Soni
quotes numbers at 2.0\,GeV, but no $\Lambda$ is given.  We use our
value for $\Lambda^{(4)}$ as well as 200\,MeV.  We calculate
$\func{B_B^{}}{m_b}$, scale it to 2.0\,GeV using $n_{\!f}$=4, and
calculate a $\hat{B}_B$ with both 4 and 5 flavors.  Since all of the
``raw'' values are close to 1.0, differences between the estimates of
the static $\hat{B}_B$ are due not to the choice of action, but to
choices in the coefficients.  See~\cite{Christensen96} for the
justification of our choice.

\end{document}